\documentclass[12pt,twocolumn,footinbib]{iopart}
\usepackage{cite}
\usepackage{times}
\usepackage{epic}
\usepackage{graphicx}

\newcommand{\be}{\begin{equation}}
\newcommand{\ee}{\end{equation}}
\newcommand{\bea}{\begin{eqnarray}}
\newcommand{\eea}{\end{eqnarray}}
\newcommand{\sub}[1]{_\mathrm{#1}}
\newcommand{\ssub}[1]{$_\mathrm{#1}$}
\begin{document}

\title[``Color-tripole ice'', conceptual generalization of ``spin ice''] {``Color-tripole ice'' as a conceptual generalization of ``spin ice''}

\author{Chia-Ren Hu}

\address{Department of Physics and Astronomy, Texas A\&M University, 4242 TAMU, College Station, TX 77843-4242, USA}

\ead{crhu@tamu.edu}

\begin{abstract}
``Spin Ice'' is an exotic type of frustrated magnet realized in ``pyrochlore'' materials Ho\ssub{2}Ti\ssub{2}O\ssub{7}, Dy\ssub{2}Ti\ssub{2}O\ssub{7}, Ho\ssub{2}Sn\ssub{2}O\ssub{7}, etc., in which magnetic atoms (spins) reside on a sublattice made of the vertices of corner-sharing tetrahedra. Each spin is Ising-like with respect to a local axis which connects the centers of two tetrahedra sharing the vertex occupied by the spin. The macroscopically degenerate ground states of these magnets obey the ``two-in two-out'' ``ice rule'' within each tetrahedron. Magnetic monopoles and anti-monopoles emerge as elementary excitations, ``fractionalizing'' the constituent magnetic dipoles. This system is also a novel type of statistical mechanical system. Here we introduce a conceptual generalization of``spin ice'' to what we shall call ``color-tripole ice'', in which three types of ``color charges'' can emerge as elementary excitations, which are Abelian approximations of the color charges introduced in high energy physics.  Two two-dimensional (2D) models are introduced first, where the color charges are found to be 1D and constrained 2D, respectively. Generalizations of these two models to 3D are then briefly discussed, In the second one the color charges are likely 3D. Pauling-type estimates of the "residual (or zero-point) entropy" are also made for these models.
\end{abstract}

%
\maketitle
\section{Introduction}
\label{sec:intro}
Frustration and fractionalization are two fundamental concepts in modern condensed matter physics. Frustration simply means the existence of competing interactions that cannot be minimized simultaneously. It could result from more than one kind of interactions present in the system, but when the interactions are all of one kind, frustration can still arise from the geometric arrangement of the constituent entities of the system (i.e., atoms, spins, etc.), it is then referred to as "geometric frustration".~\cite{geomfrus1,geomfrus2} Spin systems have offered paradigmatic examples of these concepts:~\cite{frusSpinSys-book} Spin glass is a simple example of a richly frustrated system, where ferromagnetic and antiferromagnetic bonds are randomly distributed in a spin system, leading to disordered, macroscopically degenerate ground states, and a finite residual or zero-point entropy. Antiferromagnetic Ising model on a two-dimensional triangular lattice is a simple example of geometric frustration, with also dis-ordered, macroscopically degenerate ground states and finite residual entropy, even though there is only one kind of (antiferromagnetic) interaction in the system, acting between all nearest-neighbor pairs of ``Ising spins'', which just mean quantized magnetic dipoles in strong uni-axial local fields. Fractionalization is another fundamental concept which refers to the emergence of elementary excitations which behave as parts of the elementary constituent entities of the system. A well-known one-dimensional system exhibiting fractionalization is the linear polymer trans-polyacetylene (CH)$_n$, where each electron added to the system can fractionalize into two elementary excitations, one carries the charge of the electron only, and the other carries the spin of the electron only.~\cite{SSH} This is now known as spin-charge separation. A well-known two-dimensional example of fractionalization is provided by the fractional quantum Hall effect,~\cite{FQHE} in a two-dimensional electron gas subject to a strong perpendicular magnetic field. At $1/(2n+1)$ filling of the lowest Landau level, an electron or hole added to the system can fractionalize into $(2n+1)$ elementary excitations, each of which carries $1/(2n+1)$ of the electron or hole charge (and presumably also the spin, except that in the presence of a strong magnetic field the spin is no longer a dynamic variable).

``Spin ice'' is an exotic type of magnet in which both geometric frustration and fractionalization are realized. To the knowledge of this author, it is the first three-dimensional system known to exhibit fractionalization. It occurs in pyrochlore materials Ho\ssub{2}Ti\ssub{2}O\ssub{7}, Dy\ssub{2}Ti\ssub{2}O\ssub{7}, and Ho\ssub{2}Sn\ssub{2}O\ssub{7}, etc.~\cite{spinice-expt1,spinice-expt2,spinice-expt3} In these systems magnetic atoms, with uncompensated spins, are located on a sub-lattice that is formed with the vertices of corner-sharing tetrahedra. The strong easy axis of each spin is along a line connecting the centers of two tetrahedra sharing the vertex where the spin resides, so the spin is called an Ising spin. Each tetrahedron has four spins at its four corner vertices; each of these spins can point either toward or away from the center of this tetrahedron. If a spin at one vertex points away from this center, it would be pointing toward the center of the neighboring tetrahedron that shares this vertex. If the spins have pairwise, antiferromagnetic, nearest-neighbor exchange interactions, then the four spins at the four vertices of a tetrahedron will either all point toward or away from the center of this tetrahedron, and the whole system is only doubly degenerate, corresponding to flipping all spins at the same time. But if the spins have pairwise, ferromagnetic, nearest-neighbor, exchange interactions, the four spins will obey the ``two-in, two-out'', ``ice rule'' and the system will have macroscopically degenerate ground states, and finite zero-point (or residual) entropy per spin, showing that such a system has geometric frustration,~\cite{spinice-th1,spinice-th2} but it was also shown that in the actual systems exhibiting this``spin-ice'' behavior, the magnetic dipole-dipole interactions between the spin magnetic moments are just as important, and the exchange interaction can actually be weakly anti-ferromagnetic, and the spin-ice behavior can still be obtained. The reason that the ``two-in two-out'' rule is called ice rule is because it is exactly similar to the rule governing the proton-displacement configuration in the ground state of the usual water ice: Each oxygen atom in water ice is at the center of a virtual tetrahedron with its four corners being at the half-way points between this oxygen atom and four nearest-neighbor oxygen atoms. On these four virtual line segments reside four hydrogen atoms, but only two of them are at a ``shorter distance'' to the center oxygen atom (than the corners of the tetrahedron), forming two covalent bonds with the center oxygen atom, and together, a water molecule. The remaining two hydrogen atoms are at a relatively ``longer distance'' from the center oxygen atom, (than the corners of the tetrahedron), corresponding to two hydrogen bonds, since these two hydrogen atoms are also at ``the shorter (covalent-bond) distance'' from two neighboring oxygen atoms. Thus each hydrogen atom can be viewed as to have displaced from the half-way point between the center oxygen atom and a neighboring oxygen atom, to being either closer to the center oxygen atom (the in-state of this hydrogen atom from the point of view of the center oxygen atom), or farther away from the center oxygen atom (the out-state of this hydrogen atom from the point of view of the center oxygen atom). The ground state rule of water ice is therefore also 'two-in two-out", so that each oxygen atom has exactly two hydrogen atoms covalently bounded to it. This similarity between spin-ice and water ice implies that the zero-point entropies of these two systems should be the same. The zero-point entropy of water ice has already been estimated a long time ago by the well-known theoretical chemist Linus Pauling~\cite{pauling} to be $(R/2)\ln (3/2)$ per mole of the hydrogen atoms, [or $R\ln (3/2)$ per mole of the water molecules,] where $R$ is the ideal gas constant. It was obtained using a very crude argument (see later), but this result has turned out to be quite accurate, as the zero-point entropy of the spin-ice state has already been measured,~\cite{spinice-0Tentropy-expt} and good agreement with this predicted value has been obtained. If we concentrate on the spin ice state, it is clear that if one flips an infinite chain or a closed loop of spins, all originally pointing along a single direction of the chain or loop, then the ice rule remains obeyed everywhere, and the original ground state is merely transformed into another ground state. (This is because to every tetrahedron passed through by the line or loop, there will be equal number of in-spins flipped to out-spins as that of out-spins flipped to in-spins.) But if one flips a finite chain of spins inside the system, then the ice rule will be violated at the two end tetrahedra, creating a magnetic monopole and a magnetic anti-monopole as elementary excitations in the system. These magnetic monopoles and anti-monopoles would interact with each other, and with other magnetic monopoles and anti-monopoles that happen to have also been created in the system, exactly like the usual magnetic monopoles and anti-monopoles discussed in electrodynamics books. In this sense, the system is a three-dimensional example of fractionalization: The elementary constituent entity in this system is magnetic dipole, which can now be fractionalized into a magnetic monopole and a magnetic anti-monopole,~\cite{monopole} whenever a pair of elementary excitations are created in the system. This pair creation obeys the rule that the system remains magnetic-charge neutral. Spin ice also constitutes a rich statistical mechanical system in the $(H,T)$ plane, where $H$ is an externally applied magnetic field, and $T$ is the absolute temperature. The result also depends on the direction of the applied magnetic field with respect to the crystal axes; \textit{viz.} for example, the magnetization of the system as a function of the applied field can behave very differently depending on whether the magnetic field is applied in the (1,0,0), (1,1,0), or (1,1,1) direction.~\cite{statmech}

In this work, we will present a conceptual generalization of spin ice to what we shall call ``color-tripole ice'', which also has macroscopically degenerate ground states, and non-vanishing zero-point entropy. But it will have three kinds of (so far Abelian) ``color charges'' (to be defined in Sec. II) as elementary excitations, These ``color charges'' share fundamental properties with the color charges introduced in high energy physics, and can be named red, green, and blue as well, with the property that their sum is ``colorless'' or ``color neutral'', if these color charges have ``equal strengths'' (as defined below). Note that these color charges are the conceptual generalizations of magnetic monopole, and each color charge has its corresponding anti-color charge to annihilate it, just like an anti-monopole can annihilate a monopole. Also, ``color tripole'' is a straightforward conceptual generalization of magnetic dipole, as we shall see when we define it in Sec. II. We shall then in  Secs. III and IV present two two-dimensional (2D) models of color-tripole ice, as conceptual generalizations of spin ice, which is also magnetic-dipole ice, and show that in these models, color-charged excitations can emerge, and always in color-neutral combinations, just like a magnetic monopole excitation can be created in spin ice (assumed without boundaries) if and only if a magnetic anti-monopole excitation is also created at the same time. Of course, in these models of color-tripole ice, a color charge can also be created with its anti-color charge. We find that in our first 2D model these color-charge excitations are only 1D objects. In spite of this disappointing short-coming, this model is still a very interesting statistical-mechanical system, since when a color-charged excitation moves in the system, it always has a type of ``Dirac string'' trailing behind it, which cannot be crossed by another color-charged excitation. There are three types of such Dirac strings, each oriented in a unique direction, and attached to a unique color-charged excitation. Two Dirac strings of different types cannot cross, but three Dirac strings of different types can converge at a single vertex or a single triangle (which in this 2D model is the analog of a tetrahedron in the pyrochlore spin-ice). The second 2D model is richer in behavior than the first 2D model, and is contrived to come closer to the the behavior of pyrochlore spin ice, which exists in 3D, and has 3D monopole and antimonopole excitations. But we find that even with the conception of this second 2D model, we still cannot make the color-charged excitations truely 2D objects, although the Dirac strings are no longer simple straight lines of fixed orientations, and with proper choice of the starting color-tripole crystal state, the color charge excitations created can be constrained 2D objects,  We conjecture that no 2D models of color-tripole ice can have true 2D color-charged excitations. The best we can obtain is a color-charged excitation which can move forward into any point in a wedge-shaped region of the 2D plane bounded by two semi-infinite straight lines emanating from the starting position of the color-charged excitation, or move backward into the ``opposite wedge''. (See Sec. IV below.)

In section V we will show that both 2D models of color-tripole ice can be extended to 3D models. Whereas the color-charged excitations in the first 3D model are still 1D objects (though with zigzag trajectories), we suggest that the corresponding excitations in the second 3D model is truely 3D. Thus we finally have found a satisfactory generalization of the 3D pyrochlore spin ice to 3D color-tripole ice, although in this work we have only given a very brief description of these 3D color-tripole models. In section VI we shall present Pauling-type estimates of the zero-point entropy for all four models of color-tripole ice introduced here. In section VII we present summary and conclusions, where we propose how the two 2D models of color-tripole ice might be realized via artificial fabrication. Even though all ingredients seem to be not yet already available in reality, they can likely be found or made in the future. As we see it, it will be much more difficult, but not totally impossible, to realize the two 3D models of color-tripole ice via artificial fabrication or in real materials. Even though the precise lattice structures are already determined, we still cannot see the precise entities to play the role of color tripoles, and how they can be stacked into the desired lattices, in order to realize the two 3D models of color-tripole ice. But perhaps some molecular crystals or some yet-unknown complex ordered states can realize these models. Recent advances in self-assembly of 3D binary nanocrystal superlattices also gives us some hope.~\cite{NanoxtalSuperlattice}

\section{Definitions of color charges and color tripoles used in this work}
\label{sec:def}

We begin by introducing a (in general hypothetical) two-component Coulomb charge: ${\bf q}\equiv (q_e,q_m)^{tr}$ where the superscript ``tr'' means transpose. We use the notations $q_e$ and $q_m$ for the two components,  because the electric and magnetic charges in the appropriate units (such as the Gaussian units) can already provide a realization of these two charge components, although in principle the electric and magnetic charges are surely not absolutely needed for realizing these two components. We further require (as part of the definition) that two such two-component vector charges ${\bf q}_1$ and ${\bf q}_2$, located at ${\bf r}_1$ and ${\bf r}_2$, respectively, will interact with each other via the generalized Coulomb's law:
\be
{\cal V}({\bf r}_1, {\bf r}_2) = {\bf q}_1\cdot{\bf q}_2/
|{\bf r}_1 - {\bf r}_2|\,.
\label{Coulomb}
\ee
We then define the ``red'',``green'', and ``blue'' color-charges of unit strength, $\hat{\bf q}\sub{r}$, $\hat{\bf q}\sub{g}$, and $\hat{\bf q}\sub{b}$, as simply the following three unit column vectors in the $(q_e, q_m)$ plane:
\be
\hat{\bf q}\sub{r}\equiv(1,0)^{tr}\,;\,\,\,\,\,\hat{\bf q}\sub{g}\equiv(-1/2, \sqrt{3}/2)^{tr}\,;\,\,\,\,\,\hat{\bf q}\sub{b}\equiv(-1/2, -\sqrt{3}/2)^{tr}\,,
\label{colorcharges}
\ee
which form a closed equilateral triangle in the ($q_e$, $q_m$) plane. (More generally the triangle involved does not have to be equilateral. But we shall leave that possible extension to later investigations.) It is then clear that these three unit color charges add up to zero charge in the ($q_e$, $q_m$) plane which can be referred to as colorless, or color neutral. The negative of the above three color charges are the anti-red, anti-green, and anti-blue, which can also be named cyan, magenta, and yellow, respectively.  They can annihilate the respective color charges. All of these features are similar to those of the color charges introduced in high energy physics, except that the latter ones are non-Abelian, whereas the color charges introduced here are so far Abelian. (Generalization to non-Abelian charges are potentially possible, but the realization of them in condensed matter systems will be most-likely even more difficult.)  Chin~\cite{chin} has shown a long time ago that the Abelian color charges defined above can be obtained from the non-Abelian color charges introduced in high energy physics by a well-defined Abelian approximation.

\begin{figure}[htp]
\begin{center}
\includegraphics[width=3in,angle=0,read=.pdf,type=pdf,ext=.pdf]{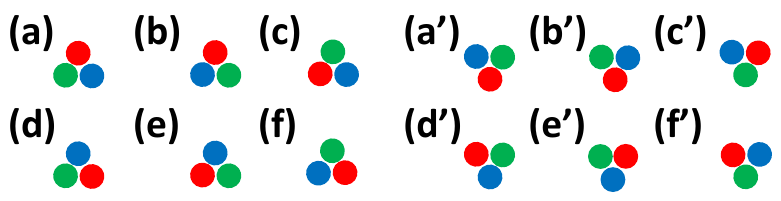}
\caption{The twelve configurations of color tripole needed to construct the two 2D models of color-tripole ice. The first six configurations, (a)-(f), are needed in both models. The next six configurations, (a')-(f'), are needed in the second model only.}
\label{fig:config}
\end{center}
\end{figure}

Next, we will define a color tripole of strength $p\sub{0}$ using a limiting process, much like a magnetic dipole of strength $p$ being defined with a limiting process as follows: A magnetic dipole has a location and a direction, so we first introduce a real-space vector of length $a$ pointing in that direction with its center located where the dipole is. We then imagine that a magnetic charge (a monopole) of strength $q\sub{M}$ is placed at the front end of that vector, and a magnetic charge of strength $-q\sub{M}$ (an anti-monopole) is placed at the tail end of that vector. If we then take the limit of letting $a\rightarrow 0$ and $q\sub{M}\rightarrow\infty$, such that the product $q\sub{M}a\rightarrow p$, we then obtain a magnetic dipole of strength $p$ pointing in the direction of the original real-space vector. Now analogously we can define a color tripole as follows: We take three color charges, one red, one green, and one blue, as defined above, but all enhanced from the basic unit color charges to strength $q\sub{0}$, and place them at the three corners of an equilateral triangle of side length $a$ in real space [at, say, ${\bf r}_1 = (0, a/\sqrt{3}, 0)$, ${\bf r}_2 = (-a/2, -a/2\sqrt{3}, 0)$, and ${\bf r}_3 = (a/2, -a/2\sqrt{3}, 0)$, respectively]. We then take the limit of letting $a\rightarrow 0$ and $q\sub{0}\rightarrow\infty$, such that the product $q\sub{0}a\rightarrow p\sub{0}$, The result is what we shall call a ``color tripole'' of strength $p\sub{0}$ in the standard configuration. It can be easily shown that this color tripole is nothing but an ``electric'' dipole and a ``magnetic'' dipole, in the sense defined above, in mutually perpendicular directions, with both dipoles having the strength $\sqrt{3}p\sub{0}/2$, so that the ``electric'' dipole is pointing toward the red charge from the center of the opposite edge of the original position triangle, and the ``magnetic'' dipole is pointing from the blue charge toward the green charge. Thus even though this color tripole does have a direction, it is not represented by a unit vector, but is rather represented by a triad (of three mutually perpendicular unit vectors) --- that is, a local coordinate system that is fixed to the original real-space triangle with three ordered vertices (where the red, green, and blue charges are placed in sequence, before the limit is taken). The interaction energy between two such color tripoles can be easily obtained by combining the interaction between the two constituent ``electric'' dipoles, and that between the two constituent ``magnetic'' dipoles. If both of the two color tripoles lie flat in the ($xy$) plane, so that their relative position vector $({\bf r}_1 - {\bf r}_2)$ also lies in this plane, then their mutual interaction takes the following simple forms: Let ($\hat{\bf e}_{x1}$, $\hat{\bf e}_{y1}$, $\hat{\bf e}_{z1}$) and ($\hat{\bf e}_{x2}$, $\hat{\bf e}_{y2}$, $\hat{\bf e}_{z2}$) denote the two triads of the two color tripoles, such that $\hat{\bf e}_{y1}$ and $\hat{\bf e}_{y2}$ ($-\hat{\bf e}_{x1}$ and $-\hat{\bf e}_{x2})$ point to the directions of the two constituent electric (magnetic) dipoles, respectively, and ($\hat{\bf e}_x$, $\hat{\bf e}_y$, $\hat{\bf e}_z$) denote the triad fixed to the lab frame. We first consider $\hat{\bf e}_{z1} = \hat{\bf e}_{z2} = \hat{\bf e}_z$.  If the orientations of ($\hat{\bf e}_{x1}$, $\hat{\bf e}_{y1}$) and ($\hat{\bf e}_{x2}$, $\hat{\bf e}_{y2}$) are obtained from ($\hat{\bf e}_x$, $\hat{\bf e}_y$) by right-handed rotations about $\hat{\bf e}_z$ by angles $\psi_1$ and $\psi_2$, respectively, and the vector $({\bf r}_1 - {\bf r}_2)$ makes an angle $\phi$ with $\hat{\bf e}_x$, then we can show that the interaction takes the simple form:
\be
{\cal V}({\bf r}_1, {\bf r}_2) = -(3/4)p_{01}p_{02}\cos(\psi_1-\psi_2)/|{\bf r}-{\bf r}'|^3\,
\label{Interaction2Rotatedtripoles}
\ee
independent of $\phi$. If in addition to rotation, one color tripole is also ``flipped over'', in the sense that its green and blue charges are swapped in positions, before the limit is taken, [corresponding to changing the signs of ($\hat{\bf e}_{x1}$, $\hat{\bf e}_{z1}$) or ($\hat{\bf e}_{x2}$, $\hat{\bf e}_{z2}$), but not both sets,] then the interaction energy becomes:
\be
{\cal V}({\bf r}_1, {\bf r}_2) =
-(3/4)p_{01}p_{02}\cos(\psi_1 + \psi_2 - 2\phi)/|{\bf r}-{\bf r}'|^3\,.
\label{InteractionOneFlippedTripole}
\ee
If both color tripoles are ``flipped over'', then the interaction is just like neither are ``flipped over''. In Sec. III, when we discuss our first 2D model of color-tripole ice, we shall need only the first six configurations of the color tripole shown in Fig.~\ref{fig:config}, (a) through (f), corresponding to fixing the orientation of the initial position triangle to be ``upward pointing'' (in the plane of the paper), and only allow the six different ways to order the three vertices, corresponding to the six elements of the symmetry group of an equilateral triangle. It would then be a generalization of an Ising spin, and will be called an Ising color tripole.

In section IV, we shall need six more configurations, corresponding to starting initially with a downward pointing position triangle (in the plane of the paper) in their definitions. [See Fig. \ref{fig:config}, (a') through (f').] The first six configurations correspond to limiting the $\psi$'s to 0$^\circ$, and $\pm 120^\circ$, plus ``flipping'' them ({\it i.e.,} rotating them about their local $\hat{\bf e}_y$ axes by $180^\circ$), whereas the next six configurations correspond to adding $180^\circ$ to the above $\psi$ angles. (The second six configurations are not parity inversions of the first six, since we require the orientation triad to be always right-handed.). Thus the interaction energies of two color tripoles in these twelve configurations can all be easily obtained. Altogether, they can be given as a $12\times 12$ matrix.

Explicit numerical study of these models will be given in a future publication. Here we concentrate on the qualitative properties of these models.

\section{The first 2D model of color-tripole ice}
\label{sec:2DModelI}

\begin{figure}[htp]
\begin{center}
\includegraphics[width=3in,angle=0,read=.pdf,type=pdf,ext=.pdf]{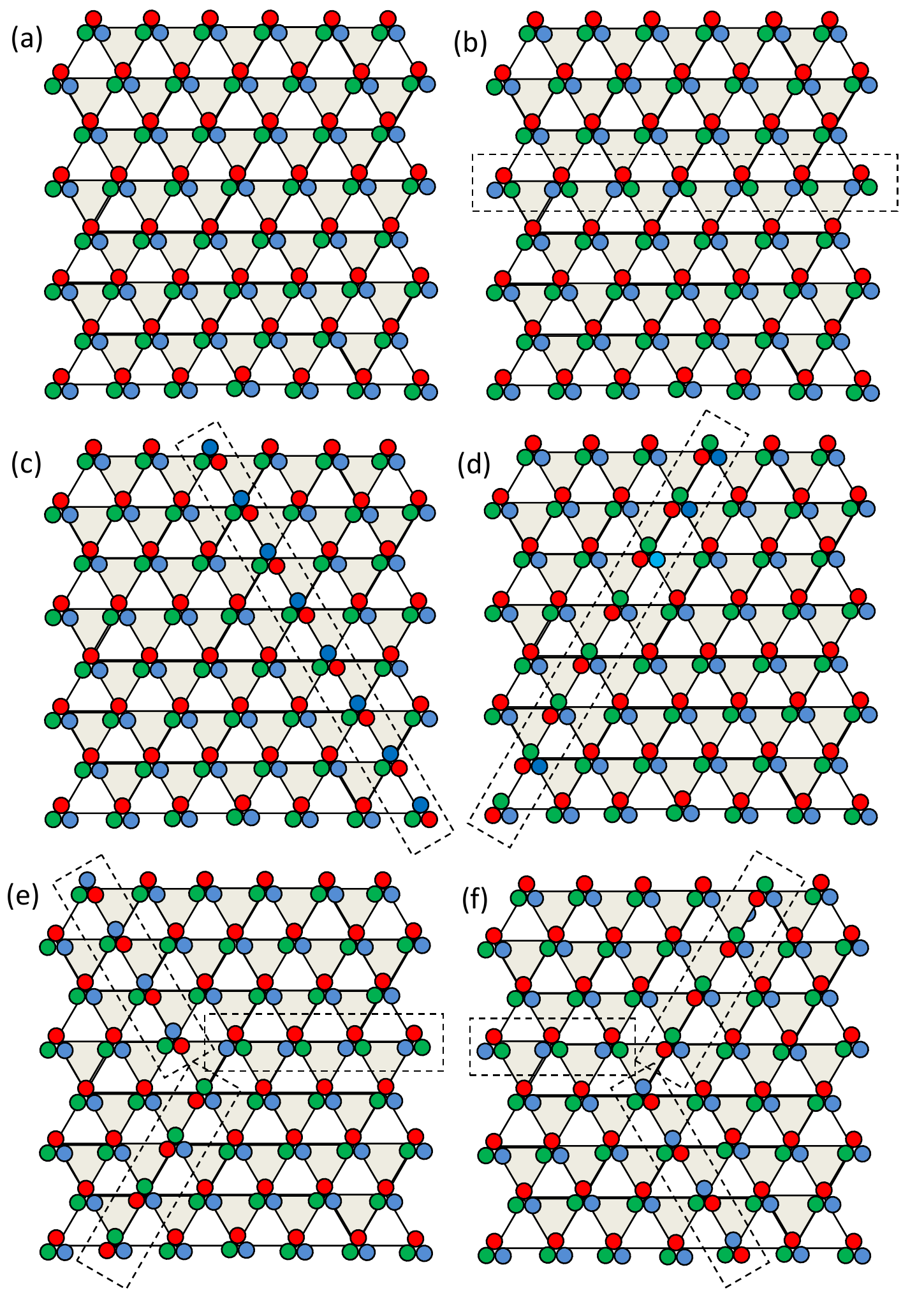}
\caption{The first two-dimensional model of color-tripole ice. (a) A color-tripole-crystal state constructed using configuration (a) in Fig.~\ref{fig:config} only. (b) A green/blue swap line introduced in the crystal state of figure (a) (marked by a dashed rectangle). (c) A red/blue swap line introduced in the crystal state of figure (a). (d) A red/green swap line introduced in the crystal state of figure (a). (e) A type-I rightward-pointing Y-intersection of three color swap lines, one of each kind. (f) A type-I leftward-pointing Y-intersection of three color swap lines, one of each kind.}
\label{fig:2DModelI}
\end{center}
\end{figure}

To construct this model we begin with a triangular lattice viewed as many downward-pointing triangles sharing vertices. Each vertex is seen to be shared by three downward-pointing triangles. Thus we place at each vertex an Ising color tripole which takes only one of the six configurations (a) - (f) in Fig.~\ref{fig:config}. That is, before the limiting process these configurations have all started with an upward pointing position triangle. Each such color tripole will then place exactly one color charge inside each of the three downward-pointing triangles of the lattice that are sharing this vertex, and each downward-pointing triangle has exactly three color charges placed in it by the three color tripoles at its three vertices. (No color charge is placed in any upward-pointing triangles of the lattice.) If all color tripoles were of configuration (a) in Fig.~\ref{fig:config}, say, we would have obtained a color-tripole-crystal state, which obeys a generalized ice rule defined below, and preserves the translational symmetry of the crystal. [See Fig.~\ref{fig:2DModelI} (a).]
However, we shall argue below that, like spin ice, the ground states of this model of color-tripole ice are macroscopically degenerate, if we assume that the interaction between any pair of color tripoles in this model is the color-tripole-tripole interaction discussed in the previous section, which is a straightforward generalization of the magnetic dipole-dipole interaction in spin ice. (We do not consider any possible generalized exchange interaction between nearest neighbor color tripoles, as in the case of spin ice, since spins have a quantum origin, and can give rise to an exchange interaction, whereas the color tripoles introduced here so far are purely classical objects.)  Thus we can use an argument similar to the one introduced by Castelnovo, Moessner, and Sondhi~\cite{monopole} to explain why spin ice in a pyrochlore lattice has macroscopically degenerate ground states. In their original argument, they propose a modest ``deformation'' of the interaction energy, viz., each magnetic dipole at any vertex of the pyrochlore lattice is replaced by a directed dumbbell of a  length equal to the distance between the centers of the two tetrahedra sharing this vertex. At the head end of this length is placed a magnetic monopole of strength equal to the strength of the magnetic dipole divided by this length, and at the tail end of this length is placed a magnetic anti-monopole of equal and opposite strength, This directed dumbbell is in the same orientation and centered at the same position as the magnetic dipole it replaces. Thus the magnetic dipole content of the spin-ice system has not been altered by this "deformation", although all magnetic higher-pole contents have been altered. But the higher-pole interactions (including the higher-pole-dipole interactions) all vanish with distance faster than the dipole-dipole interaction. Thus the dominant, longest-range interaction energy of the system is not affected by this ``deformation''. Castelnovo et al. further introduced a self-energy term at the center of each tetrahedron, proportional to the square of the net magnetic charge there, in order to mimic the main effect of these higher-pole interactions (and possibly also a non-dominant nearest-neighbor exchange interaction). In this way the original interaction energies are successfully mimicked up to corrections small everywhere and vanish with distance at least as fast as $1/r_{i,j}^5$, which is the distance dependence of the quadrapole-quadrapole interaction. With this ``deformation'' it becomes clear why the ground states of the pyrochlore spin ice should satisfy the ice-rule: It is so that the net magnetic charge deposited at the center of each tetrahedron by the relevant dumbbells is exactly zero, thus reducing the total interaction energy calculated in this ``deformation'' approximation to zero, which is clearly the minimum energy if compared with all those due to all possible configurations with a net monopole charge distributions in the system that are overall (i.e., globally) neutral. This is because such configurations would correspond to separating monopoles from anti-monopoles that are together in the ground states, and then place them in some distribution. Clearly such separation energy is positive no matter how one does it. Now this argument can be straightforwardly generalized to the present model of color-tripole ice: We can also introduce a moderate deformation to the interactions in the system, by replacing each color tripole at any lattice site by what is the definition of the color tripole before the limit is taken: viz., an upward-pointing equilateral triangle with three different color charges placed at its three vertices. We shall let the size of the equilateral triangle be such that the color charges become exactly placed at the centers of the three downward-pointing triangles sharing this vertex in the lattice. If the distance between the nearest neighbor vertices of the lattice is $a$, then the center-to-center distance we just referred to is also equal to $a$. Thus to keep the same color-tripole content everywhere in the lattice we shall let the strength of the three color charges be all equal to $p_0/a$, where $p_0$ is the strength of the original color tripoles being replaced. We shall name the object replacing a color tripole as a ``trumbbell'',~\cite{footnote-1} because it is a straightforward generalizaion of the dumbbell introduced in Ref.~\cite{monopole}. A dumbbell has two end points where a monopole and an anti-monopole are placed. A trumbbell has three end points (forming an equilateral triangle in the real space) where three color charges are placed --- one red. one green, and one blue, all of the same strength. They are positioned like the three end points of a Y, and a circle going through them is exactly equally divided into three identical arcs. That is, they are exactly 120$^\circ$ apart on the circle. The three color charges form an upward-pointing equilateral triangle, of exactly the same size as the upward-pointing triangles of the lattice, but the three color charges on this trumbbell are positioned exactly at the centers of three downward-pointing triangles of the lattice that share the vertex where the original color tripole replaced by the trumbbell is positioned. With such a ``deformation'' approximation, the total interaction energy of the system becomes a sum of the interactions between pairs of {\it net} color charges (analogous to the net magnetic charges at the centers of the tetrahedra in a pyrochlore spin ice after the ``deformation'' approximation), located at the centers of the downward-pointing triangles of the lattice, possibly with self-energy corrections just like the original argument presented in Ref.~\cite{monopole}
The minimum value of this total interaction energy is again zero, corresponding to no net color charge deposited at the center of any downward-pointing triangle of the lattice. This is the generalized ice rule for defining the ground states of this model --- color neutrality within each downward-pointing triangle, or exactly one red charge, one green charge, and one blue charge, all of the same strength, placed at the center of each downward-pointing triangle of the lattice by the three trumbbells located at the three vertices of this triangle. That this zero total interaction energy is the minimum value because any other net color charge distribution in the lattice, which must be overall color neutral, must result from taking apart originally locally neutralized color charge distributions. This generalized ice rule naturally leads to macroscopically degenerate ground states, just like the pyrochlore spin ice studied in Ref.~\cite{monopole}, since the two arguments are closely parallel. In Sec.~\ref{sec:0ptentropy} we shall give a Pauling-like estimate of the zero-point entropy of this model color-tripole ice. We shall see that it has a finite value per color tripole, much like pyrochlore spin ice.

\begin{figure}[htp]
\begin{center}
\includegraphics[width=3in,angle=0,read=.pdf,type=pdf,ext=.pdf]{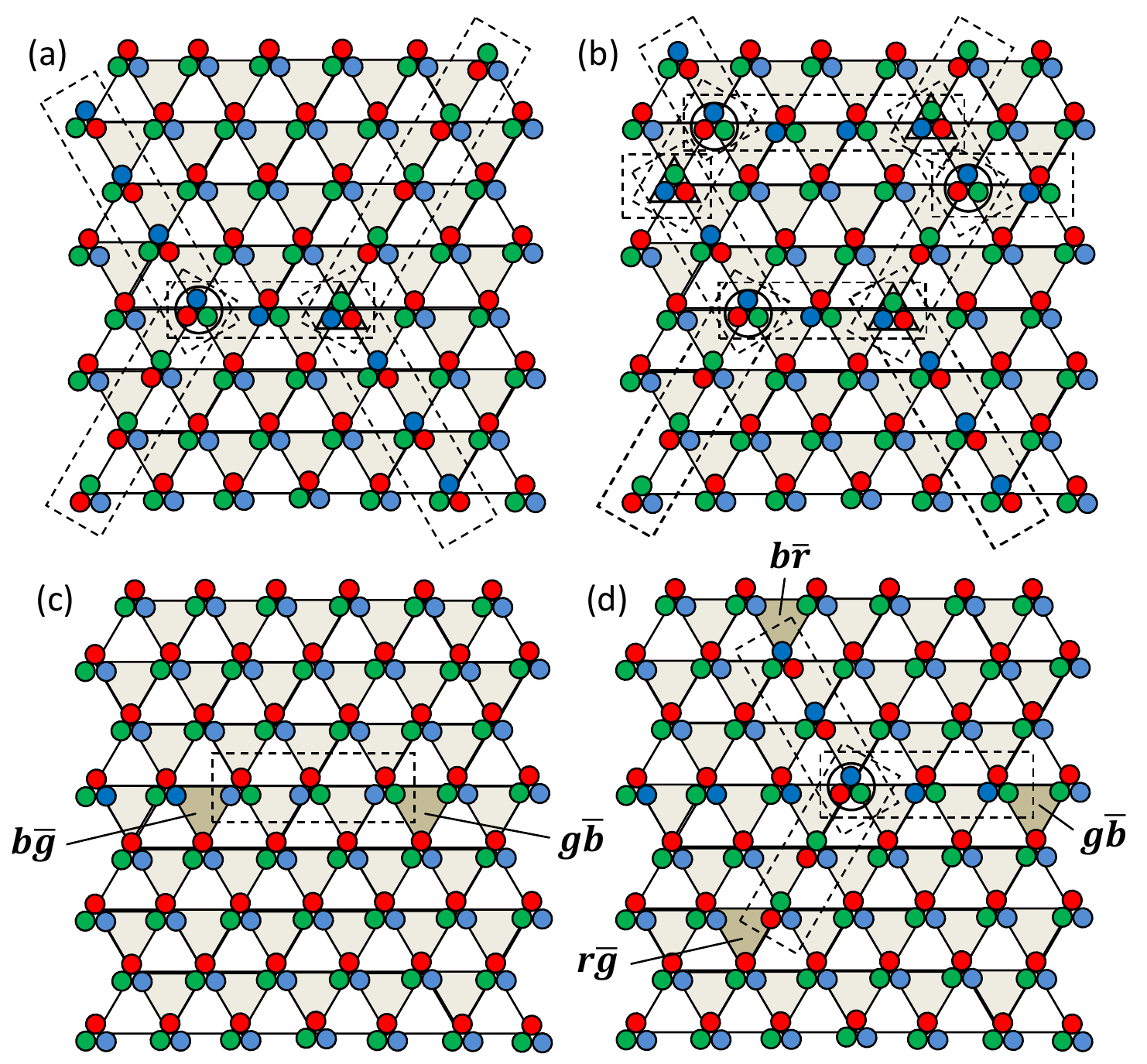}
\caption{The first two-dimensional model of color-tripole ice --- continuation. (a) Two type-II Y-intersections of color-swap lines introduced in the color-tripole crystal state of Fig.~\ref{fig:2DModelI} (a), the left one is rightward pointing, and the right one, leftward pointing. (b) An example of a ground state containing six Y-intersections, all type-II in this example. (c)  At the two ending downward-triangles of a finite-length color-swap line (shaded darker grey), two color-charged elementary excitations appear, which can combine to color neutrality. (d) At the three outerwardly-ending downward-triangles of a finite-sized Y-intersection (shaded darker grey), three color-charged elementary excitations appear that can combine to color neutrality.}
\label{fig:2DModelI-2}
\end{center}
\end{figure}

Starting with the color-tripole crystal state introduced already, which obeys the generalized ice rule, (or any of the other five color-tripole crystal state which can be obtained from this crystal state by applying one of the five non-trivial symmetry operations of an equilateral triangle to all color tripoles in the lattice at the same time,) we can see how other ground states can be generated, which also obey the generalized ice rule. First, we can swap green and blue charges within each of a {\it horizontal} straight line of color tripoles that extends to infinity at both ends, assuming that the crystal is also infinitely large. [See Fig.~\ref{fig:2DModelI} (b).] We shall call this line a ``green/blue swap line''. We can also have a ``red/blue swap line'', but it must be an infinitely-long straight line making $+120^\circ$ (or, equivalently, $-60^\circ$, since the line is not directional) with horizontal, --- see Fig.~\ref{fig:2DModelI} (c), or a ``red/green swap line'', but it must be an infinitely-long straight line making $-120^\circ$ (or, equivalently, $+60^\circ$) with the horizontal --- see Fig.~\ref{fig:2DModelI} (d). These color-swap lines can be viewed as three types of generalized ``Dirac strings''. There are no line energy densities associated with them within the ``deformation'' approximation. Two such color-swap lines cannot cross. However, three such color-swap lines, one of each kind, can all terminate at the three vertices of a single downward triangle, and the generalized ice rule will still be satisfied everywhere. [See Fig.~\ref{fig:2DModelI} (e) and (f).] We shall call them ``type-I Y-intersections''. The first one is ``rightward-pointing'', and the second one, ``leftward-pointing''. Three such color-swap lines can also all terminate at the same lattice site --- called type-II Y-intersections, except that at this intersection site a new configuration must be assigned to the color tripole in order to maintain the generalized ice rule everywhere: If the Y-intersection is rightward-pointing, then at the intersection vertex the color tripole should be assigned the configuration (e) in Fig.~\ref{fig:config} [See the left Y-intersection in Fig.~\ref{fig:2DModelI-2} (a).] On the other hand, if the Y-intersection is lefttward-pointing, then at the intersection vertex the color tripole should be assigned the configuration (f) in Fig.~\ref{fig:config} [See the right intersection in Fig.~\ref{fig:2DModelI-2} (a).] In general, many such intersections can appear when assigning Ising color tripoles to all vertices of the lattice, without violating the generalized ice rule, thus all giving zero total energy in the ``deformation'' approximation. [See Fig.~\ref{fig:2DModelI-2} (b), where only type-II Y-intersections are used, which does not have to be the case.] Thus the ground states can be highly degenerate. The degree of degeneracy, and hence the zero-point entropy of this color-tripole ice, will be investigated in Sec.~\ref{sec:0ptentropy} using a Pauling-like argument.

If any of the color-swap lines has a finite length and terminates at two downward-pointing triangles of the lattice, we would have created two color-charged elementary excitations in the lattice, one having the anti-color of the other. That is, they can combine to color neutrality. For example, if a green/blue swap line is terminated on both ends at two downward-pointing triangles, then its right-end downward triangle will have a net color charge of green-plus-anti-blue, or in short, $g\bar b$, and its left-end downward triangle will have a net color charge of blue-plus-anti-green, or in short, $b\bar g$. They are anti-colors of each other, so they can combine to color neutrality. [See Fig.~\ref{fig:2DModelI-2} (c).] Similarly, terminating the other two color-swap lines can generate elementary excitations of colors red-plus-anti-blue ($r\bar b$), blue-plus-anti-red ($b\bar r$), red-plus-anti-green ($r\bar g$), and green-plus-anti-red ($g\bar r$).  If one cut out a finite section in the middle of an infinitely-long color-swap line, one can also generate two color-charged elementary excitations that can combine to color neutrality. On the other hand, if one terminates the three ``semi-infinite color-swap lines'' of any Y-intersection, at three downward triangles of the lattice, one would have created three color-charged elementary excitations that can combine to color neutrality. For example, in Fig.~\ref{fig:2DModelI-2} (d), the three end downward-triangles have net color charges of $b\bar r$, $g\bar b$, and $r\bar g$, respectively. They clearly can combine to color neutrality. Note that these color-charged elementary excitations are only one-dimensional objects, even though the model is two-dimensional, since they must stay at the end of a generalized Dirac string, which can change its length, but not its direction, nor can it bend. This point may be considered a short-coming of this model, since in the three-dimensional pyrochlore spin ice, the monopole/anti-monopole excitations are also three dimensional objects. This problem can be traced to the fact that in the case of pyrochlore spin ice, there are two monopoles and two anti-monopoles deposited inside each tetrahedron, and yet in the present model of color-tripole ice, there is only one red, one blue, and one greeen color charge deposited inside each downward-pointing triangle of the lattice. This gives no freedom for a Dirac string to exit a downward triangle after entering it. In this sense this model is not ``perfect''. Nevertheless, this 2D model of color-tripole ice will still be a very interesting and novel statistical-mechanical system, not only because the long-range Coulomb-like interaction between the emergent color-charged elementary excitations, but also because the three types of Dirac strings, which must trail behind the color-charged elementary excitations, cannot cross each other. Thus the only way for a color-charged elementary excitation to move through a Dirac string that is in the way, is to cut apart that Dirac string, thus creating two new color-charged elementary excitations. Since such creation costs finite energy, there must exists an onset temperature before such a process can take place. We thus can foresee a complex temperature dependence in the equilibrium properties of this system. Furthermore, These ``two-component'' color charges can see a ``two-component field'', which can be electric and magnetic fields in the particular way to realize these color charges discussed in Sec.~\ref{sec:intro}. Thus the field-dependent properties of this system must be also very rich, especially since each field component can still be applied in a different orientation in relation to the lattice. In Sec.~\ref{sec:SumCon} we shall speculate on how our proposed models might be realized.

\section{The second two-dimensional model of color-tripole ice}
\label{sec:2DModelII}

\begin{figure}[htp]
\begin{center}
\includegraphics[width=3in,angle=0,read=.pdf,type=pdf,ext=.pdf]{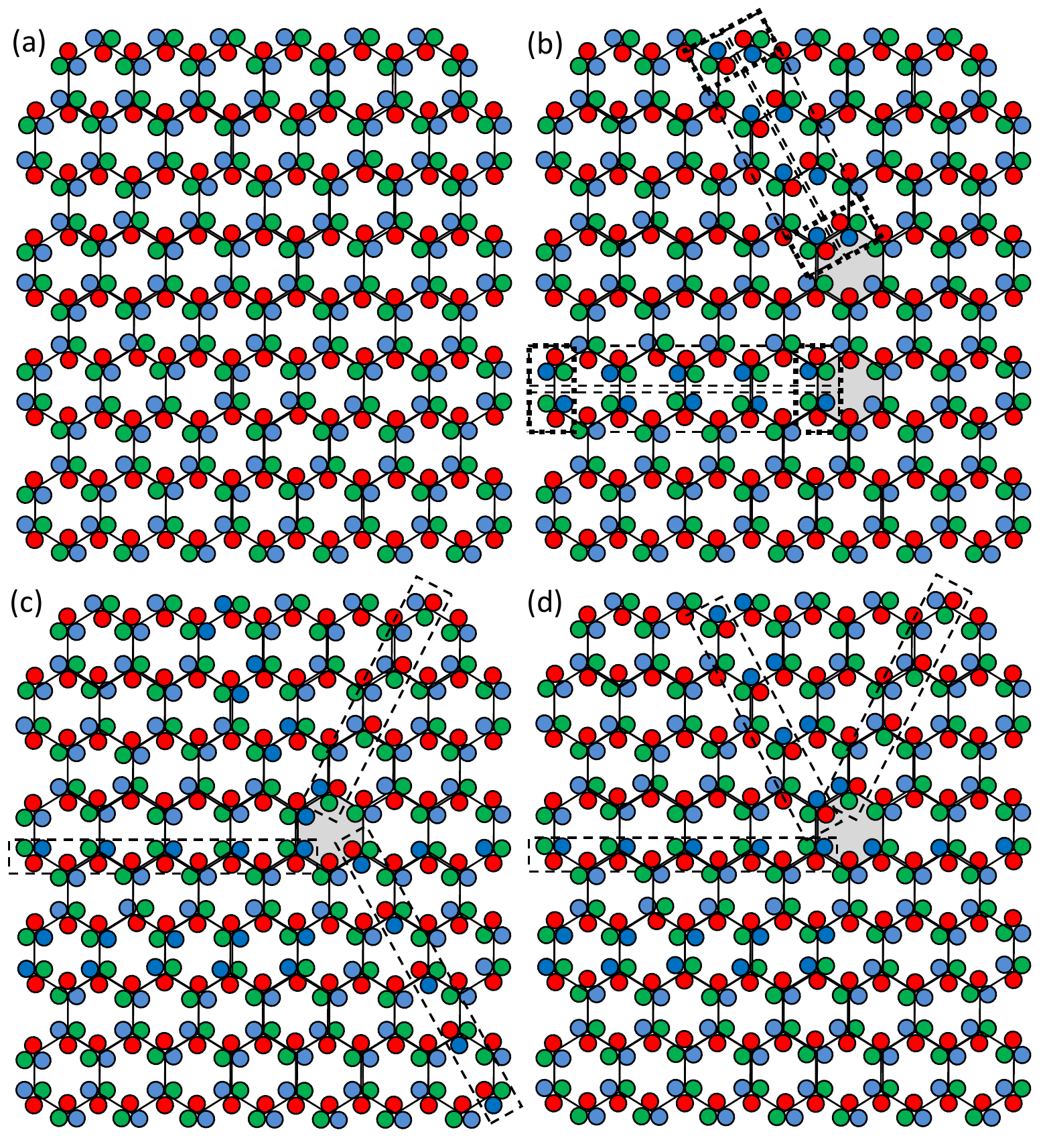}
\caption{The second two-dimensional model of color-tripole ice. (a) One of the 36 color-tripole crystal states that obey the generalized ice rule for color-tripole ice, and preserve the translational symmetries of the lattice. It is made of configurations (a) and (a') of Fig.~\ref{fig:config}. (b) An example of a ``ground state'' containing two ``collapsed loops'', each of which is made of two finite-and-equal-length color-swap lines that end on the same hexagon on its one end (shaded grey) and on another hexagon on its other end (which is outside the plotted region). (c) A ``usual kind'' of type-I Y-intersection (at the hexagon marked grey) of three color-swap lines making 120$^\circ$ with each other (enclosed in dotted rectangles). Note that in this case the three color-swap lines all reside on one type of lattice sites (here type A). (d) An ``unusual kind'' of type-I Y-intersection of three color-swap lines which do not make 120$^\circ$ with each other. We see that both types of lattice sites are now involved in the three color-swap lines in this case.}
\label{fig:2DModelII}
\end{center}
\end{figure}

In this model we begin with a hexagonal lattice. We note that each vertex in this lattice is shared by three hexagons. However, we also note that all vertices of this lattice can be classified into two types: A type-A vertex and its three nearest neighbor vertices form an upright Y. At each type-A vertex one must put a color tripole which is among the configurations (a) through (f) in fig.~\ref{fig:config}, so that it can place a color charge inside each of the three hexagons sharing this vertex. A type-B vertex and its three nearest neighbor vertices form an inverted Y. At each type-B vertex one must put a color tripole which is among the configurations (a') through (f') in Fig.~\ref{fig:config}, in order to achieve the same purpose. In Fig.~\ref{fig:2DModelII} (a) we have shown a color-tripole crystal state in this lattice which is constructed with configurations (a) and (a') only.

Note that without breaking the translational symmetries of this lattice, one can form $6\times 6 = 36$ crystal states in this lattice, by using different combinations of configurations in the two types of lattice sites. Starting with any one of these crystal states, one can form other ground states in a way similar to the first 2D Model. For example, if one starts with the crystal state in Fig.~\ref{fig:2DModelII} (a), one can also introduce three types of straight color-swap lines as in the first 2D Model, except that in this new model the color-swap lines can now reside on either a line of type-A lattice sites, or a line of type-B lattice sites. In addition, this model has more exotic possibilities: A color-swap line can now make a 180$^\circ$ turn at a hexagon of the lattice, changing from residing on type-A lattice sites to residing on type-B lattice sites, or vice versa. But if a color-swap line makes two such 180$^\circ$ turns, it would only return to the original line that it has started in, rather than effectively move in 2D. The result is a collapsed loop made of two finite-and-equal-length color-swap lines of the same type [{\it c.f.,} Fig.~\ref{fig:2DModelII} (b)],  but residing on two different types of lattice sites. This system can also have many more kinds of Y-intersections. Two examples of type-I Y-intersections are given in Fig.~\ref{fig:2DModelII} (c) and (d); the first one is the ``usual kind'', and the second one is an ``unusual kind''. One can also construct a ``usual'' kind of type-II Y-intersections, but we shall omit their details here. (Note that an ``unusual kind'' of type-II Y-intersection does not exist!) Clearly, the statistical-mechanical properties of this model will be even more complex, but the color-charged elementary excitations, although they can now move on folded lines and even doubly-folded lines ({\it i.e.,} collapsed loops), are clearly still not truely 2D objects.
\begin{figure}[htp]
\begin{center}
\includegraphics[width=3in,angle=0,read=.pdf,type=pdf,ext=.pdf]{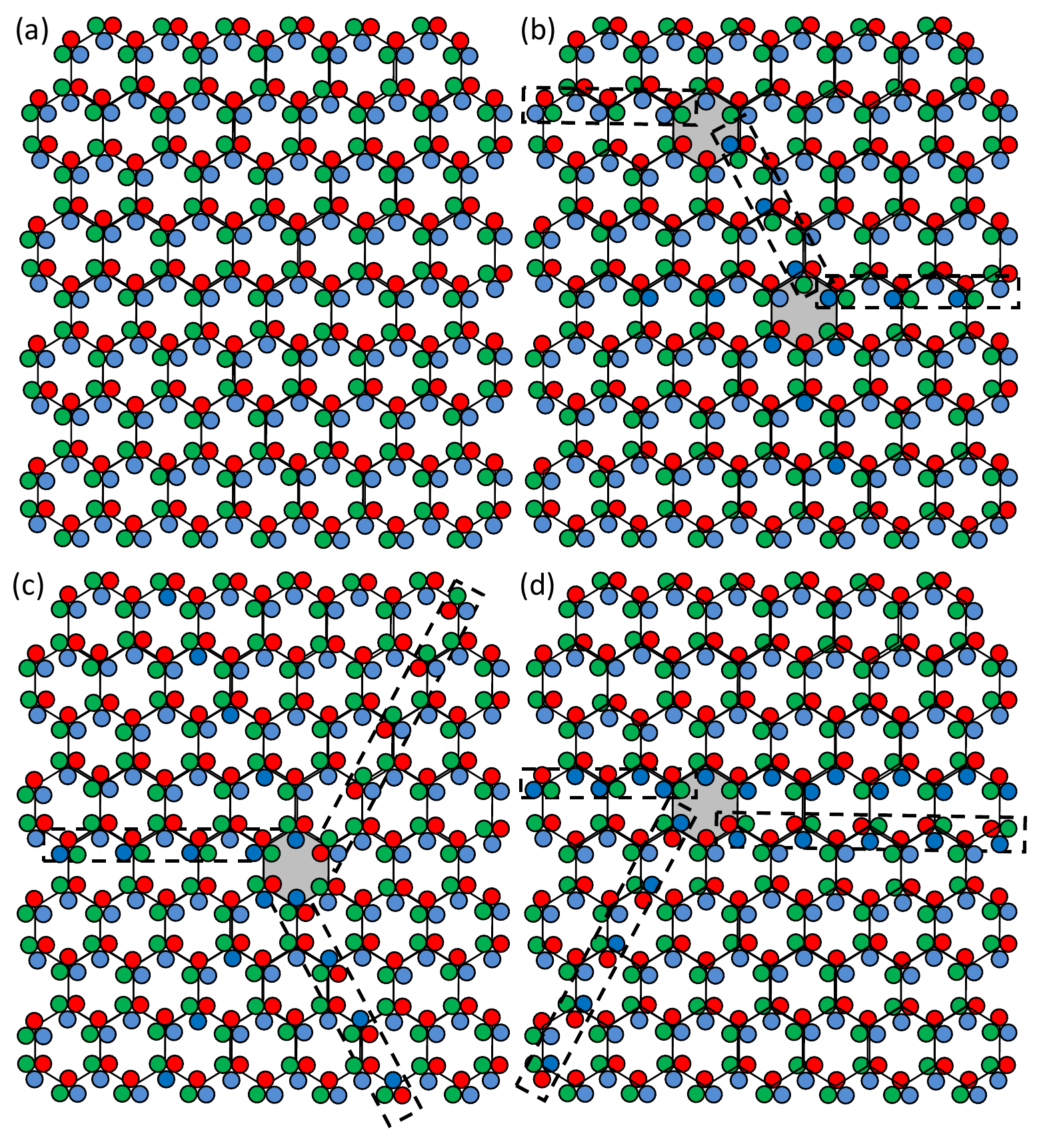}
\caption{The second two-dimensional model of color-tripole ice --- continuation. (a) A different one of the 36 color-tripole crystal states that obey the generalized ice rule for color-tripole ice, and preserve the translational symmetries of the lattice. It is made of configurations (a) and (e') of Fig.~\ref{fig:config}. (b) An example of a ground state containing a twice-turned blue/green color-swap line (at the two grey hexagons) inserted into the crystal state in (a). (c) A ``usual kind'' of type-I Y-intersection of three color-swap lines which make 120$^\circ$ with each other, inserted into the crystal state of (a). In this case the three color-swap lines all reside on one type of lattice sites. (d) An ``unusual kind'' of type-I Y-intersection, where the three color-swap lines do not make 120$^\circ$ with each other. In this case both types of lattice sites are involved in the three color-swap lines.}
\label{fig:2DModelII-2}
\end{center}
\end{figure}

Now suppose we start with a different color-tripole crystal state, which is constructed with configurations (a) and (e') of Fig.~\ref{fig:config}, say, as shown in Fig~\ref{fig:2DModelII-2} (a), then a rightward-extending green/blue swap line can make a 60$^\circ$ turn to the right (but not any degrees to the left) by changing from residing on type-A vertices to type-B vertices [c.f., Fig.~\ref{fig:2DModelII-2} (b)], But after that it can no longer make more turns to the right. Instead, it can only either continue forward, or make a 60$^\circ$ turn to the left, and return to the very direction it has started in [c.f., Fig.~\ref{fig:2DModelII-2} (b)]. (If one starts with a blue/red color swap line residing on type-A lattice sites and extending toward lower right, it can make a 120$^\circ$ turn to the left, and becomes residing on type-B lattice sites, and extending toward upper right. If it makes another turn, it also returns to the original direction. (We omit presenting a figure for this case, and for the behavior of a red/green color swap line which involves $\pm$60$^\circ$ turns.) This system can also have the ``unusual kind'' of Y-intersections [c.f., Fig.~\ref{fig:2DModelII-2} (d)], beside the ``usual kind'' [c.f., Fig.~\ref{fig:2DModelII-2} (c), where the three color-swap lines are 120$^\circ$ apart, as is the case when all color-swap lines reside on one type of lattice sites only.]. This second model will also have macroscopic number of ground states, to be analyzed in Sec.~\ref{sec:0ptentropy}. When a color-charged elementary excitation is created in this model, by terminating some color-swap line at some hexagon, we can see that it is no longer constrained to move on a straight line, but can cover a wedge-shaped area in the 2D plane, bounded by two semi-infinite straight lines emanating from any starting point of the elementary excitation, or it can go backward to cover the opposite wedge. This is because a color-swap line can now make a turn of a certian angle (to the right by 60$^\circ$ or to the left by 120$^\circ$, depending on which type of color swap line it is, assuming that it has originally resided on type-I lattice sites. If it has originally resided on type-II lattice sites, it will be to the left by 60$^\circ$ or to the right by 120$^\circ$.) Thus these elementary excitations are still not truely 2D objects, which should be able to go almost anywhere in a 2D plane. Thus even though this second 2D model is contrived to reproduce an important feature of the pyrochlore spin ice, viz., each tetrahedron in that lattice has {\it two} magnetic charges of each kind deposited at its center by the dumbbell replacements, In the present second 2D model, we also have two color charges of each kind deposited at the center of each hexagon by the trumbbell replacements, unlike the first model, where only one color charge of each kind is deposited at the center of a downward triangle. But because the vertices of the hexagonal lattice are clearly separated into two types, We are still not able to achieve our goal --- obtaining truely 2D color-charged elementary excitations in a 2D model, although they can now at least do a constrained 2D motion. Below we discuss only briefly how these 2D models can both be generalized to 3D. We think that the second 3D model can have truely 3D color-charged excitations, However we plan to make detailed discussion of these 3D models in a future publication.

\section{Generalization of the first 2D color-tripole-ice model to 3D}
\label{sec:3DModel-I}

\begin{figure}[htp]
\begin{center}
\includegraphics[width=3in,angle=0,read=.pdf,type=pdf,ext=.pdf]{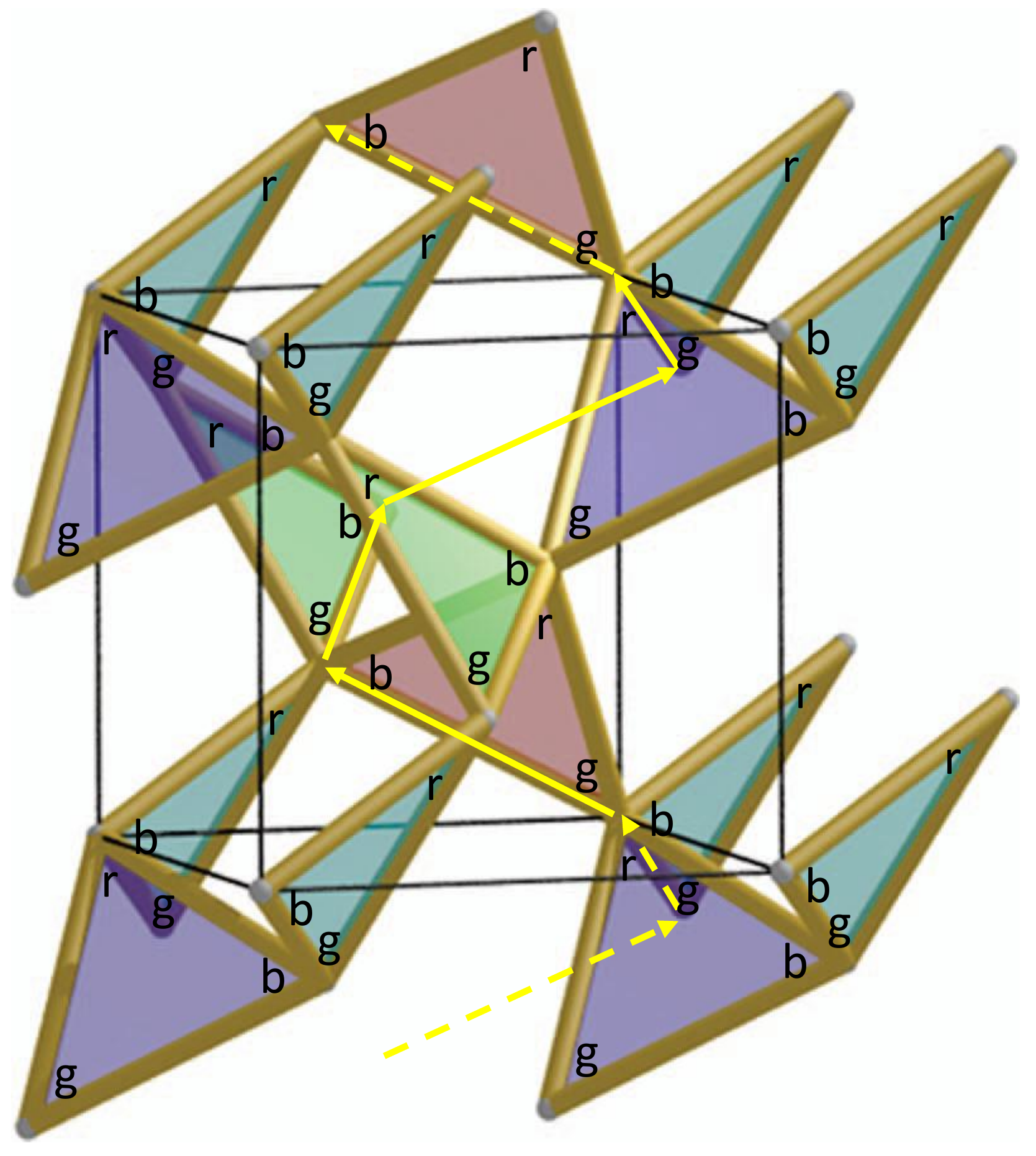}
\caption{The trillium lattice cubic unit cell (from Fig. 1 of Ref.~\cite{trillium-1}), together with a color-tripole crystal state defined on it, in the trumbbell approximation, in which the color tripole at the center of each equilateral triangle of the lattice is replaced by three color charges, one red (r), one green (g), and one blue (b), which are deposited on the three vertices of the triangle, as shown. At each vertex (or lattice site) there are three color charges deposited, by the three trumbbells located at the centers of the three equilateral triangles sharing this vertex. The chain of yellow arrows indicates the path for a blue-green swap line in this crystal state. (The dashed parts are the periodic extension of the solid part.)}
\label{fig:trillium}
\end{center}
\end{figure}

For this purpose we begin with the so-called ``trillium lattice''.~\cite{trillium-1,trillium-2,trillium-3} [See Fig.~\ref{fig:trillium}.]
In this lattice each lattice point is a shared vertex of three equilateral triangles, each of which is formed by three lattice points at its three vertices. There are four orientations of these equilateral triangles, which are related to each other like the four (1,1,1) orientations of a cube (but not the cube shown). To construct the 3D generalization of our first 2D model of color-tripole ice, we must put a color tripole at the center of each equilateral triangle of the lattice, in the orientation of that triangle. It would then have only six options to go in. In the trumbbell approximation, this color tripole will be replaced by three color charges, one red, one green, one blue, each of which is deposited at one vertex of the triangle. Each vertex will then have three color charges deposited at its location, originating from the three color tripoles at the centers of the three triangles sharing this vertex, These three color charges, in the ground state, will combine to color neutrality --- which is the generalized ice rule defined in this work for color-tripole ice. A color-tripole crystal state is defined in Fig.~\ref{fig:trillium}, given in the trumbbell approximation. A potential path for a green-blue color swap line is also indicated in this figure as a chain of yellow arrows. (But no color swaps have been done yet.) Such lines can keep the generalized ice rule intact. Clearly, red-green color swap lines and blue-red color swap lines can also be defined and introduced into this crystal state, without violating the generalized ice rule.  Type I and II Y-intersections can also be defined and introduced into this crystal state. Thus color-neutral combinations of pairs or triplets of color-charged elementary excitations can also be generated in this 3D model by terminating the color-swap lines. But since the color-swap lines have fixed shapes and orientations with no path-branching possible, the color-charged excitations generated in this 3D model of color-tripole ice are still 1D objects only, although they do not move in simple straight lines. For example, elementary excitations with color charges $b\bar g$ and $g\bar b$ can only move on paths such as the one indicated by the chain of yellow arrows in Fig.~\ref{fig:trillium} or in the reversed path. Such color-swap lines again can not cross each other, nor can a color-charged excitation cross a color-swap line, and these excitations cannot take alternative paths locally to avoid each other either! Numerical study of the statistical mechanical properties of this model will be deferred until when realization of this model becomes at least with good probability possible.

\section{Generalization of the second 2D color-tripole-ice model to 3D}
\label{sec:3DModel-II}

For this purpose we begin with simply the ``pyrochlore lattice'' already mentioned in association with spin ice. We note that one view of this lattice is that each vertex is shared by two tetrahedra, but another view of this lattice is that each vertex is shared by six equilateral triangles, three of which are the faces of one tetraheron, and the other three of which are the faces of the other tetrahedron. Thus to construct the 3D generalization of our second model of color-tripole ice, we must put a color tripole at the center of each equilateral triangle of this lattice, in the orientation of that triangle. In the trumbbell approximation, this color tripole will again be replaced by three color charges, one red, one green, one blue, each of which is deposited at a separate vertex of the triangle. Each vertex will then have six color charges deposited at this site, originating from the six trumbbells replacing the six color tripoles located at the centers of the six triangles sharing this vertex, These six color charges, in the ground state, will again combine to color neutrality, thus obeying the generalized ice rule defined in this work for color-tripole ice. These six color charges must then be {\it two} red, {\it two} green, and {\it two} blue. Thus once a color-swap line enters a vertex, it now has two choices on how it can exit the vertex --- in much analogy with the pyrochlore spin ice, where once a spin-flip line enters  a tetrahedron, to preserve the ice rule it can have two choices on how it can exit the tetrahedron, because each tetrahedron has four vertices where spins reside, with {\it two} of them pointing into, and the remaining {\it two} of them pointing out of the tetrahedron.

We shall defer to a future work to introduce even a color-tripole crystal state in this model, and to give explicit examples of the color-swap lines. These can be combined with a numerical study of the statistical mechanical properties of this model, which is likely very rich in properties. We only note here that the color-charged elementary excitations, which can emerge in this model, will most likely be 3D, much like the monopole excitations in pyrochlore spin ice, because of the properties of any color swap line described in the previous paragraph.

There may be even better lattices to realize this 3D model of color-tripole ice, where each vertex is shared by six (or nine? twelve? fifteen? ...) equilateral triangles that are more evenly distributed around this vertex, instead of being grouped into those of two (or more?) tetrahedra. But so far we have not found them. The readers with more experiences with exotic 3D lattices might be able to find them. Or perhaps mathematicians might be able to prove the non-existence of any of such cases.

\section{Zero-point entropy}
\label{sec:0ptentropy}

We first review the Pauling estimate of the zero-point entropy of pyrochlore spin ice. For this purpose the interactions between the spins are totally neglected, and are simply replaced by the ice rule. If the system has $N$ spins, then it will have $2^N$ possible configurations. What fraction of them obeys the ice rule? This fraction is estimated by treating all tetrahedra in the pyrochlore lattice as if they are totally independent of each other. Each tetrahedron has $2^4 = 16$ configurations, and the number of configurations which satisfy the ice rule is $4!/2!2! = 6$. Hence the desired fraction in one tetrahefron is $6/16 = 3/8$. The total fraction of ground states in the whole system is therefore estimated to be $(3/8)^{N/2}$ since each spin is shared by two tetrahedra, and yet each tetrahedron has four spins at its four vertices, so there must be $N/2$ tetrahedra in the system. Multiplying this fraction by the total number of configurations $2^N$, we obtain the desired estimate of the total number of ground states in this system to be $(3/2)^{N/2}$, giving the total zero-point entropy of this system to be $k_B*(N/2)\ln(3/2)$, or $(R/2)\ln(3/2) = 0.20273 R$ if $N = N_A$, for one mole of spins. (Here $k_B$, $R$ and $N_A$ are Boltzmann constant, ideal gas constant, and the Avogadro number, respectively.) This result is also the total zero-point entropy of water ice, per mole of hydrogen atoms, or per half-mole of water molecules.

We now apply similar reasonings in order to estimate the zero-point entropy of the four models of color-tripole ice introduced here. For this purpose we begin by assuming that there are $N = N_A$ color tripoles residing on $N_A$ vertices. (In the two 3D models we would be talking about the vertices of the ``dual lattices'', which are the face-centers of the original lattices introduced.) In the first 2D model, there are also $N_A$ downward equilateral triangles in the lattice, each having three vertices, but each vertex is shared by three equilateral triangles. For each downward equilateral triangle in the lattice, the fraction of configurations that satisfy the generalized ice rule is equal to $(3!\times 2^3)/6^3 = 2/9$. Thus we can estimate the number of ground states to be $6^{N_A}\times (2/9)^{N_A} = (4/3)^{N_A}$, giving the molar zero-point entropy to be $R\ln(4/3) = 0.28768 R$.

Next, we consider the 3D generalization of this model. Referring to the figure in Fig~\ref{fig:trillium}, we see that each cubic cell has four lattice points and four equilateral triangles associated with it. Thus the number of lattice points of the trillium lattice is equal to the number of its ``dual lattice'' points. So there are also $N_A$ vertices which correspond to equilateral triangles in the dual lattice. To each such triangle, we again find the fraction of configurations that satisfy the generalized ice rule to be equal to 2/9. Thus we can again estimate the the number of ground states to be $(4/3)^{N_A}$, giving the molar zero-point entropy to be again $R\ln(4/3) = 0.28768 R$.

Next, we consider the second 2D model. In this model, we have each vertex shared by three "hexagons", and each hexagon having six vertices. Thus to $N_A$ vertices must correspond $N_A/2$ hexagons. Within each of these "hexagons" the fraction of configurations obeying the generalized ice rule is $[(6!/2!2!2!)\times 2^6]/6^6 = 10/81$. Thus we can estimate the number of ground states to be $6^{N_A}\times (10/81)^{N_A/2} = (40/9)^{N_A/2}$, giving the molar zero-point entropy to be $R/2\ln(40/9) = 0.74583 R$.

Finally, we consider the 3D generalization of the second 2D model. Here we must consider the ``dual lattice'' of the pyrochlore lattice, with each ``dual-lattice'' point being the center of an equilateral-triangle face of a tetrahedron in the pyrochlore lattice. Since each equilateral triangle has three original lattice points as its vertices, and each original lattice point is shared by six equilateral triangles, we see that $N_A$ ``dual-lattice'' points correspond to $N_A/2$ original lattice points. Each ``dual-lattice'' point is also shared by three ``hexagons'', except that each of these ``hexagons'' is not flat, but is heavily corrugated, since its six vertices are the centers of six equilateral triangles that share the same lattice point of the original pyrochlore lattice. three of these six triangles are the faces of one tetrahedron, and the other three are the faces of another tetrahedron. The said lattice point of the pyrochlore lattice is shared by these two tetrahedra. The rest of the analysis is exactly the same as that for the second 2D model, as the number of these ``corrugated hexagons'' is just the number of lattice points of the original pyrochlore lattice. We thus conclude that the molar zero-point entropy of this 3D generalization of the second 2D model of color-tripole ice is also $R/2\ln(40/9) = 0.74583 R$.

Note that the infinite-temperature molar entropies of these four models of color-tripole ice are all equal to $R\ln 6 = 1.79176 R$. Thus the zero-point entropies of the second 2D model and its 3D generalization are already about 42\% of their infinite-temperature entropies, showing that the ground-state degeneracies of these two models are very high! On the other hand, the zero-point entropies of the first 2D model and its 3D generalization are only about 42\% higher than that of the pyrochlore spin ice. They are only about 16\% of the infinite-temperature entropies, whereas the zero-point entropy of pyrochlore spin ice is already about 29\% of its infinite-temperature entropy. This is because the molar infinite-temperature entropy of spin ice is only $R\ln 2 = 0.69315 R$, which is much smaller than that of the color-tripole ice, $R\ln 6 = 1.79176 R$.

Our estimate of the zero-point entropies presented above are all based on a generalization of the original Pauling estimate. Whether this approximation is reasonably accurate or much cruder for the color-tripole ice than for the spin ice must await an essentially exact numerical study of of the color-tripole-ice models.

\section{Summary and Conclusion}
\label{sec:SumCon}

In this work, we have introduced a conceptual generalization of pyrochlore spin ice to what we have called ``color-tripole ice''. It has macroscopically degenerate ground states and non-vanishing molar zero-point entropy just like spin ice. However, unlike spin ice, which has magnetic monopole/antimonopole elementary excitations, fractionalizing magnetic dipoles which are the elementary building blocks of that system, in the case of color-tripole ice, what are fractionalized are what we called ``color tripoles'', resulting in three kinds of color-charged elementary excitations, and their corresponding anti-color-charged elementary excitations. These color charges are analogous to the color charges introduced in high energy physics, except that the former are Abelian so far, whereas the latter are well-known to be non-Abelian. Further generalization to non-Abelian color charges may be possible, but the chance to realize them in condensed matter systems (natural or artificially fabricated) would be even more difficult, so we have not yet given it much thought so far. Instead, we have concentrated on finding ways to realize the Abelian models introduced here. Thus before we conclude this section, we would like to give some general suggestions to this goal. We have proposed in this work two two-dimensional (2D) models and then discussed their possible three-dimensional (3D) generalizations. We deemed that realizing either of the two 3D models will be much more difficult than realizing the corresponding 2D models. So below we will only discuss the possible ways to realize the two 2D models by artificial fabrication. To arrange entities in a triangular or hexagonal lattice can presumably be done in the laboratories by a number of methods, including optical and e-beam lithography, masks and etching, and even by the use of a scanning tunneling tip to put atoms in atomic-scale patterns. Thus we need only concentrate on the realization of the individual color tripoles which should be arranged in the prescribed locations in the said lattices as described in the two 2D models. For this purpose we must first realize the two-component, fictitious, ``vector Coulomb charges'' introduced in Sec.~\ref{sec:def}. As we have already explained in that section, one way to realize these ``vector Coulomb charges'' is to use electric and magnetic charges in appropriate relative strengths. Then after introducing an appropriate unit system the generalized Coulomb's law given in Eq.~\ref{Coulomb} between these ``vector Coulomb charges'' can be realized. We have then shown that an ideal color tripole is equivalent to an electric dipole and a magnetic dipole, in appropriate relative strengths, and locked in mutually perpendicular orientations. The oriented plane containing these two dipoles is then the oriented plane of the color tripole. [The orientation of a color tripole is described by a triad of three mutually perpendicular unit vectors ($\hat{\bf e}_1$, $\hat{\bf e}_2$, $\hat{\bf e}_3$), and the said oriented plane is spanned by the first two unit vectors.] For example, with the color charges defined in Eq.~\ref{colorcharges} (a), a unit color tripole in the first configuration as shown in Fig.~\ref{fig:config} (a) --- which has been defined as the standard configuration --- in the ($\hat{\bf e}_1$, $\hat{\bf e}_2$) plane, is just equivalent to an electric dipole of strength $\sqrt{3}/2$ in the said unit pointing in the $\hat{e}_2$ direction, and a magnetic dipole of strength $\sqrt{3}/2$ in the said unit pointing in the $-\hat{e}_1$ direction. The other configurations we need in the two 2D models are just this standard configuration rotated in the plane by $\pm 120^{\circ}$, $180^{\circ}$, and $\pm 300^{\circ}$, possibly with in addition a mirror reflection with respect to the $\hat e_2$ axis, inverting $\hat e_1$ to $-\hat e_1$. (To keep the triad right-handed, $\hat e_3$ must also be  inverted.) These procedures give a total of 12 configurations, as depicted Fig.~\ref{fig:config}. A color tripole that is limited to either the first six or the second six configurations of Fig.~\ref{fig:config} has been called an Ising color tripole. To realize them by artificial fabrication we can possibly use magnetoelectric multiferroic materials~\cite{Magnetoelectrics}, which exhibit both spontaneous electric polarization and spontaneous magnetization.~\cite{multiferroics} Since they have different transition temperatures one might be able to adjust their relative strengths by adjusting the temperature. But this is not a good way to realize the color tripole since then one does not have the 2D color-tripole-ice models fabricated at all temperatures. So it is better that the two ordered states occur at the same transition temperature, or, more easily, both at very high temperatures, so that at the temperature range of interest, both orders are practically temperature independent. Then a way to adjust their relative strengths is to use two coupled layers of controlled thicknesses, one ferroelectric and one ferromagnetic. Or one can combine a multiferroic layer with a ferroelectric or ferromagnetic layer of carefully chosen thicknesses, if only these layers are also coupled, so that the ferroelectric polarizations or ferromagnetic magnetizations in the two or more layers can be always aligned. Two more conditions must still be met: (1) The permanent electric dipole moment $\bf P$ and the permanent magnetic dipole moment $\bf M$ in the (possibly composite) material must be locked in mutually perpendicular orientations, (2) In the plane of the color-tripole there should be strong three-fold easy axes so that the color tripole so generated can only be in the first six configurations of Fig.~\ref{fig:config} in its intrinsic ($\hat e_1$, $\hat e_2$) plane. This constraint might be achieved by shaping the material to an equilateral triangular slab, in order to acquire shape anisotropy. Alternatively, it appears that three tantalum atoms in close proximity to form an equilateral triangle might have both permanent electric and permanent magnetic dipole moments.~\cite{3TantalumAtoms}, --- See also, Ref.~\cite{3SodiumAtoms} --- but it may be difficult to adjust their relative strength to what is required here. Also, so far we do not yet know whether their $\bf{P}$ and $\bf{M}$ prefer to stay in the plane of the atomic triangle and to be always locked in mutually perpendicular orientations. Once they are so then the three-fold anisotropy requirement will be automatically fulfilled in this case. However, it may be that the permanent electric moment can exist in the three- or more-atom cluster only if the cluster has a geometric asymmetry, ({\it i.e.,} a deviation from perfect equilateral-triangle symmetry). If so, one must worry whether the asymmetry can still exist when the structure is fabricated on  a substrate, and whether the asymmetry can rotate freely in the tripole-plane by $\pm 120^{\circ}$, and do mirror reflection about its intrinsic ${\hat e}_2$ axis. Both conditions are required for the realization of a 2D color-tripole ice in artificial fabrication. This idea might be generalized to other kinds of atoms, and to more than three atoms (four, six, nine, ten, etc.), with a three-fold symmetric arrangement in the basal plane. In fact, if three atoms of one kind can generate a magnetic moment, and three atoms of another kind can generate an electric moment, then combining them into a six-atom configuration may let one generate both. By adjusting the relative sizes of the two three-atom clusters may let one adjust the relative strengths of the two moments. Adding 
another atom at the center may let one further adjust this relative strength. Even the relative orientation of the two moments may be adjustable by changing the relative orientation of the two tri-atom clusters. Furthermore, it may be possible to use artificial atoms in place of real atoms,
or even to combine artificial atoms with real atoms to achieve this purpose. Thus the potential to realize a color tripole along this route may be high. So far our proposal to realize a color tripole is still very imprecise. We shall investigate it further and report our findings in a future publication.

There may exist various generalizations (extensions) of the idea presented here. At first sight, it may appear that instead of having three distinct colors that can combine to color neutrality, we can also have five, or seven, or even larger prime numbers of, distinct colors that can combine to color neutrality that are still based on two-component vector Coulomb charges. One can simply replace the equilateral triangle used in the definition of the three fundamental color charges presented in Sec.~\ref{sec:def} by an equilateral pentagon or heptagon, etc. Actually, it is not so simple. When the angle between the defining vectors of two different color charges in the $(q_e, q_m)$ plane are smaller than 90$^{\circ}$, they will repel rather than attract, and color-neutral combinations of color charges may prefer to be separated rather than annihilated.  
Can one start with three-or-more-component vector Coulomb charges in defining color charges? We do not yet know for sure. But the realization of them in atomic or condensed matter systems will be much more difficult, so we have not yet given them serious consideration. Another possible generalization is to employ a non-equilateral triangle in the definitions of the three color charges. It likely will lead to the equivalence of a generalized ``color tripole'' with an electric dipole and a magnetic dipole of unequal strengths in the unit system we have described. It would then make the realization of a color-tripole much easier. Further investigations will be needed to nail down the properties of the resultant color-tripole ice if it can exist.

Finite temperature properties of the color-tripole-ice models including possible phase transitions in the absence or presence of external fields coupled to the two components of the vector Coulomb charges will be investigated in a future publication. Such systems will likely have very  rich and novel behavior.

\end{document}